# Companhias aéreas são todas iguais? A convergência dos modelos de negócios no transporte aéreo

Renan P. de Oliveira
Alessandro V. M. Oliveira⁺

Instituto Tecnológico de Aeronáutica, São José dos Campos, Brasil

⁺ Autor correspondente. Instituto Tecnológico de Aeronáutica. Praça Marechal Eduardo Gomes, 50. 12.280-250 - São José dos Campos, SP - Brasil.
E-mail: alessandro@ita.br.

*Resumo*: Este estudo discute a literatura quanto à convergência de modelos de negócios de companhias aéreas no transporte aéreo brasileiro, com foco na formação de redes de voos. Primeiramente, analisa os fatores determinantes dos padrões de formação de redes dos modelos de negócios "fundamentais" (arquétipos) das companhias aéreas nos primeiros anos após a desregulamentação do setor. Em seguida, discorre sobre como os modelos de negócios das companhias brasileiras se assemelham a esses padrões. A literatura destaca convergências entre as estratégias de formação de rede das companhias de serviço completo em relação às companhias de baixo custo mais antigas, além de redirecionamentos de modelos de negócios após fusões e aquisições.

*Palavras-chave*: transporte aéreo, companhias aéreas, modelos de negócios.

## I. Introdução

A questão da convergência de modelos de negócios no setor aéreo tem ganhado destaque nos últimos anos, com diversos exemplos sendo amplamente divulgados pela mídia especializada, especialmente com foco nas operações nos céus americanos e europeus. De um lado, companhias de baixo custo, referidas daqui em diante de acordo com o jargão da área, pelo termo em inglês low-cost carriers (ou LCCs), começaram a se direcionar de forma cada vez mais evidente ao segmento de passageiros de negócios, aumentando a complexidade dos seus serviços oferecidos durante o processo (devido especialmente ao crescente alcance de seus produtos a segmentos díspares do mercado). Companhias provendo o "serviço-completo", referidas como full-service carriers (FSCs), em um outro extremo, começaram a adotar um conjunto de práticas tradicionalmente associadas as suas rivais LCCs de forma a tornarem seus modelos mais flexíveis (com medidas como reduções de custos, oferta desagregada de serviços e iniciativas de reestruturação de suas redes). Assim, estes dois modelos têm delineado e aproximado um modelo de negócios aparentemente unificado, um meio termo "híbrido" de ambos, integrando ideias de preços mais em conta e diferenciação de serviços, correndo atrás dos mesmos (segmentos de) clientes com estratégias cada vez mais similares.

Não se restringindo só à mídia, mas penetrando também o meio acadêmico, o fenômeno tem atraído um número crescente de publicações. Em particular, evidências empíricas de LCCs e FSCs se afastando de suas propostas originais de modelos de negócios estão disponíveis sob uma variedade de perspectivas, quer seja (re)configurações de redes, frequências de voos, programas de lealdade e a provisão (ou não) de comodidades a bordo (com exemplos de trabalhos associados a estes temas incluindo Mason e Morrison, 2008; Daft e Albers, 2013; Lohmann e Koo, 2013; Jean e Lohmann, 2016; Klophaus et al., 2012; Fageda et al., 2015; Henrickson e Wilson, 2016). Klophaus et al. (2012), por exemplo, fornecem evidências de que uma porção de companhias dentre aquelas que desviam do modelo idealizado (o arquétipo) LCC operacionalizam tal desvio por meio de suas escolhas de aeroportos (não se restringindo somente a aeroportos secundários) e suas estratégias de formação de rede (adotando combinações de sistemas hub-and-spoke e ponto a ponto), enquanto os resultados de Fageda et al. (2015) sugerem que tais desvios são primariamente manifestos sob os aspectos de desagregação de serviços – em especial no que tange o preço final das passagens – e, similarmente, operações ponto a ponto.

Estes últimos resultados são de particular relevância para a proposta deste trabalho, uma vez que fornecem uma ponte entre modelos de negócios e as redes de companhias aéreas. Segundo Holloway (2008), uma rede de voos de uma companhia aérea é uma manifestação explícita de seu comportamento estratégico, definindo sua identidade de marca e influenciando suas receitas e custos. Além disso – discorre o autor –, a rede de voos pode ser uma fonte de vantagem competitiva, que os concorrentes precisam considerar ao decidir enfrentá-la ou evitá-la. E também pode oferecer alguma proteção contra flutuações econômicas, especialmente quando alguns mercados são afetados mais do que outros.

Além da mídia e do meio acadêmico, outros grupos podem ter particular interesse nessa questão, em especial autoridades reguladoras, visto que LCCs desempenha(ra)m um papel essencial na expansão extraordinária do transporte aéreo a nível mundial nos últimos 25 anos e há expectativas de que continuem a fazê-lo.

Com essas considerações, algumas questões vêm naturalmente à tona: quão conformes os modelos de negócios de companhias aéreas brasileiras tem se mantido àqueles modelos arquetípicos tradicionais, em especial ao modelo LCC? A que extensão a convergência ou divergência tem ocorrido (ou já ocorreu) entre companhias e suas rivais? Para qual modelo de negócios tal convergência (caso tenha de fato ocorrido) se direcionou? Tais perguntas serão endereçadas neste trabalho, entretanto, primeiramente, uma visão panorâmica do ambiente de modelo de negócios no Brasil será apresentada.

## II. Panorama Histórico

Aqui nós consideraremos as quatro principais companhias aéreas em operação no Brasil antes da pandemia, mais precisamente, as companhias Gol, LATAM, Azul e Avianca – esta última tendo suas operações finalizadas em meados de 2019. A companhia Gol iniciou suas operações como uma LCC em 2001, passando por uma mudança significativa em seu modelo de negócios em 2007 após sua aquisição da "companhia aérea de bandeira" do Brasil, a Varig. Desde então, a companhia tem empenhado esforços no reestabelecimento de sua imagem como uma LCC. Até 2018, a companhia tem se mantido posicionada como a companhia aérea brasileira com o menor "custo por assento-quilometro disponível", uma medida de eficiência econômica amplamente encontrada na literatura de aviação, em particular sob seu nome (e, principalmente, sigla) em inglês: Cost per Available Seat Kilometer (CASK). A companhia aérea LATAM (anteriormente conhecida por TAM), por outro lado, começou suas operações como uma empresa regional de pequena escala, impulsionada pela liberalização do mercado de transporte aéreo no país em 1998 e se tornou a maior FSC do Brasil (com a falência de sua rival Varig sendo um fator contribuinte a este título). Contrariamente ao caso da aquisição da Varig pela Gol, a fusão da TAM com a companhia chilena LAN não alterou significativamente a orientação de seu modelo de negócios.

Similar à trajetória da LATAM, a companhia Avianca surgiu como um reposicionamento de marca da companhia regional OceanAir em 2010, com traços de um modelo de negócios FSC. Por fim, a companhia Azul é aqui considerada, tendo começado suas operações no aeroporto secundário de Campinas/SP e se tornado rapidamente a principal LCC do país. Este rótulo, no entanto, viria a ser contestado após sua aquisição da companhia regional Trip em 2012. Nos tempos atuais, a Azul se mantem como a companhia brasileira com o maior preço por passageiro-quilometro (medida denominada na literatura de aviação como "yield"), embora, ao mesmo tempo tenha o menor custo operacional por viagem.

### III. Convergência e Divergência de Modelos de Negócios

Nesta parte do texto, após a contextualização do mercado brasileiro, serão discutidos estudos nos quais pesquisadores procuraram entender sob quais aspectos os modelos de negócios de companhias aéreas estariam convergindo ou divergindo. Desta literatura, chamamos a atenção aos trabalhos de Mason e Morrison (2008) que, estudando o mercado europeu notaram que desvios de um modelo de negócios puramente LCC se mostraram menos lucrativos para a companhia aérea envolvida nos mesmos; o trabalho de Lohmann e Koo (2013), cujos resultados sugerem que empresas americanas envolvidas em uma fusão convergem para o modelo FSC; e o trabalho de Daft e Albers (2015) que, considerando empresas alemãs, também encontram evidências de convergência de modelos para o modelo FSC, com resultados adicionais indicando que LCCs tendem a se diferenciar mais entre elas do que FSCs.

Além disso, Klophaus et al. (2012) encontram evidências que desvios do modelo arquetípico LCC são manifestados principalmente por meio da escolha de aeroportos e estratégias de configuração de rede companhias aéreas com resultados similares sendo encontrados por Fageda et al. (2015), indicando, por meio de uma abordagem econométrica, que desagregação de serviços nos preços de passagens e desvios de operações ponto a ponto são as maiores fontes de distorções. Por fim, citamos os trabalhos de Henrickson e Wilson (2016), que ao investigar a indústria do transporte aéreo americana, também por meio de uma abordagem econométrica, sugerem que convergências ocorrem entre LCCs e FSCs, particularmente para o modelo FSC. Em suma, resultados da literatura apontam para a ocorrência de convergência de modelos de negócios tanto nos EUA quanto na Europa, particularmente com companhias aéreas se aproximando do modelo FSC. Com essas considerações, fica claro o caráter dinâmico dos modelos de negócios de companhias aéreas.

Como se pode perceber, as pesquisas mais recentes têm se preocupado em explorar as mudanças de modelos de negócios de companhias aéreas, em especial em mercados economicamente mais maduros, como EUA e Europa, onde os limites tradicionalmente aceitos que separam os modelos de negócios clássicos entre si tem sido obscurecidos. Apesar de inúmeros esforços terem sido despendidos para o entendimento da evolução dos modelos de negócios de companhias aéreas, foram poucos os estudos econométricos realizados nesta literatura. Em particular, a questão da decomposição de um modelo de negócios de uma companhia como uma mistura de modelos clássicos variante ao longo do tempo não configura um tema ainda explorado. Desta forma, um estudo nacional foi elaborado para identificar como os modelos de negócios vigentes de companhias aéreas brasileiras poderiam ser decompostos, bem como as influencias que fusões e aquisições bem como a competição de rivais poderiam influenciar a ocorrência de distorções na orientação dos modelos destas companhias.

### IV. Estratégias das Companhias Aéreas Brasileiras

No contexto brasileiro, o estudo de Oliveira, Oliveira e Lohmann (2019, 2021) abordou a questão da convergência de modelos de negócios de companhias aéreas. Modelos econométricos foram usados na investigação científica das mudanças nos modelos de negócios das companhias aeras brasileiras Azul, Avianca, Gol e LATAM, usando dados de janeiro de 2000 a dezembro de 2018. Os dados de janeiro de 2000 a dezembro de 2005 foram empregados na calibração dos arquétipos de modelos de negócios (os modelos de negócios clássicos), nomeadamente, os modelos LCC, FSC e RGC. As companhias usadas para sua calibração foram a Gol e as (então denominadas) TAM e OceanAir, respectivamente. O modelo usado para a formação dos padrões de configuração de rede dos arquétipos, o modelo do primeiro estágio, foi baseado na literatura científica existente acerca do tema de entrada de companhias aéreas em mercados, contendo variáveis referentes à distância, um termo de distância ao quadrado para capturar um efeito "amortecedor" para distâncias muito longas, a maior renda per capita entre as cidades de origem e destino de uma rota, a fim de considerar a influência econômica das mesmas, e uma variável "rede", medindo o maior número de cidades servidas, entre as cidades de origem e destino, partindo ou chegando das mesmas. Esta variável é usada para medir um fator "hub" para os aeroportos das redes das empresas. Notamos que alguns controles adicionais são considerados no estudo e o leitor interessado é convidado a consultar o material original para maiores informações.

Partindo para os resultados deste primeiro estágio, ilustrados na Tabela 1, notamos que várias similaridades podem ser observadas a respeito da direção da influência que os fatores exercem sobre a presença de um arquétipo de modelos de negócios em uma rota. A maior diferença que pode ser notada é referente às intensidades destas influências. Adotamos a convenção que a cada 0,3 pontos no coeficiente, um novo sinal (positivo ou negativo, a depender da direção do efeito) é ilustrado na entrada correspondente da tabela. Chamamos a atenção para a convenção usada de que um coeficiente ligeiramente maior que 0, porém ainda assim estatisticamente significante, receberá um sinal (positivo ou negativo) na tabela. Os coeficientes de distância são positivos, um resultado interpretado em vista de rotas mais longas terem uma menor competição por parte de modais alternativos de transporte. No entanto, notamos que o arquétipo LCC obteve um coeficiente para o termo de amortecimento (a distância ao quadrado) mais pronunciado. A tendência de todos os modelos de servir mercados com passageiros (em média) mais abastados também parece ser uma unanimidade. Por fim, a variável rede parece surtir um efeito equivalente em todos os arquétipos considerados, indicando a existência de aeroportos centrais nas redes de todas elas (embora estes aeroportos não necessitem ser os mesmos).

**Tabela 1 - Determinantes de Redes de Arquétipos**

| Características | (1) FSC | (2) LCC | (3) RGC |
|---|---|---|---|
| Distância | ++ | +++ | ++++++ |
| Distância$^2$ | - | - | ------- |
| Renda per capita | + | + | ++ |
| Rede | + | + | + |

Fonte: de Oliveira, Oliveira e Lohmann (2019). Cada 0,3 pontos no coeficiente estimado pelos autores, equivale a um novo sinal (positivo ou negativo).

Voltando-nos aos resultados do segundo estágio, nossos principais resultados, podemos perceber que, de acordo com a Tabela 2, o modelo de negócios da Azul no começo de 2006 pode ser descrito como majoritariamente LCC com traços de regional e com uma tendência a evitar o modelo FSC. Suas tendências temporais, no entanto, sugerem que ao longo de sua trajetória a companhia foi se alinhando cada vez mais com o modelo FSC e se distanciando dos modelos LCC e regional. Sua fusão com a Trip, como esperado, proporcionou um aumento na sua parcela regional, acompanhado de um aumento na sua parcela LCC e uma redução expressiva do modelo FSC.

Com relação aos resultados relativos a Gol, a companhia pode ser descrita em 2006 como majoritariamente LCC com traços FSC e com um certo afastamento do modelo regional. Com o tempo, no entanto, a companhia passa a se afastar do modelo FSC e se aproximar do modelo regional, aumentando, no processo, ainda mais sua parcela LCC. Sua aquisição da Varig proporcionou um aumento em sua parcela FSC acompanhado de uma redução em sua parcela LCC, enquanto sua parcela regional não obteve resultados estatisticamente significantes nesse quesito.

O modelo da Avianca apresenta resultados que indicam uma mistura dos três arquétipos no começo da base utilizada na pesquisa (2006), com tendências temporais de redução de sua parcela regional – como esperado após seu reposicionamento no mercado – e um aumento de suas parcelas FSC e LCC. Por fim, resultados da (então denominada) TAM indicam que a companhia era extremamente alinhada ao modelo FSC e evitava os modelos LCC e regional em 2006, com uma tendência temporal de redução do modelo FSC em detrimento de um maior alinhamento com os outros dois arquétipos. Resultados sugerem que sua fusão com a LAN, no entanto, teve um efeito de intensificar sua parcela FSC e reduzir as parcelas LCC e RGC.

Por fim, considerando os efeitos dos diferentes pares de companhias aéreas, notamos como que suas presenças (e ausências) aparentam estar correlacionadas positivamente – um indicativo da convergência entre suas redes –, com exceção do caso da Avianca e da Azul, cujos resultados sugerem a existência de uma divergência (embora moderada). Em particular, chamamos a atenção para as correlações mais intensas entre a LATAM e a Gol e a Avianca e a Gol, melhor ilustradas no material original.

**Tabela 2 - Decomposição dos modelos**

| Determinates | (1) Azul | (2) Gol | (1) Avianca | (4) LATAM |
|---|---|---|---|---|
| FSC | - | ++ | + | ++++++ |
| LCC | ++++ | ++++ | + | -- |
| RGC | + | - | + | - |
| FSC × tendência | + | - | + | + |
| LCC × tendência | - | + | + | + |
| RGC × tendência | - | + | - | + |
| FSC × fusão/aquisição | --- | + | —— | + |
| LCC × fusão/aquisição | + | - | —— | —— |
| RGC × fusão/aquisição | + | NS | —— | - |

Fonte: de Oliveira, Oliveira e Lohmann (2019). Cada 0,3 pontos no coeficiente estimado pelos autores, equivale a um novo sinal (positivo ou negativo).

**Tabela 3 – Influência das rivais**

| | (1) Azul | (2) Gol | (3) Avianca | (4) LATAM |
|---|---|---|---|---|
| (1) Azul | —— | —— | —— | —— |
| (2) Gol | + | —— | —— | —— |
| (3) Avianca | - | + | —— | —— |
| (4) LATAM | + | ++ | + | —— |

Fonte: de Oliveira, Oliveira e Lohmann (2019). Cada 0,3 pontos no coeficiente estimado pelos autores, equivale a um novo sinal (positivo ou negativo).

## V. CONCLUSÕES

Seriam todas as companhias aéreas iguais? Como discutido neste trabalho, as companhias aéreas de diversos mercados ao redor do mundo têm se tornado cada vez mais similares no que diz respeito aos serviços oferecidos por elas, devido a estas, no final das contas, buscarem atender a mesma gama de clientes – com desejos e necessidades, por vezes, díspares. No entanto, seguir exatamente o caminho que outras companhias seguem pode, muitas vezes, dificultar a sobrevivência financeiras de companhias que não conseguirem fazer frente à eficiência de rivais de modelos similares. Buscar um nicho, onde poucos atuam pode trazer uma certa estabilidade para a companhia aérea, um exemplo disso sendo o caso da "guinada" observada para a orientação do modelo de negócios da Azul em direção ao seguimento regional: muitos destes mercados são rotas nas quais a companhia se mantem até hoje livre de concorrência.

Apesar destas considerações, é importante ter em mente que o número de nichos em um mercado é limitado. Como expresso por Jacobi et al. (2014) "*não basta ter um nicho no mercado, é preciso ter um mercado no nicho*". Alguns nichos não são interessantes de serem explorados de um ponto de vista econômico, uma situação que acaba limitando o número de direções que uma companhia aérea pode decidir seguir. Os estudos nacionais o qual este trabalho tem por base (de Oliveira, Oliveira e Lohmann (2019, 2021), investigaram os efeitos de Fusões e Aquisições (F&As) e da concorrência sobre a orientação e a evolução de modelos de negócios de companhias aéreas, chegando à conclusão que convergências de estratégias de formação de rede mais acentuadas ocorreram entre as duas FSCs com relação à LCC mais antiga do mercado. Além disso, redirecionamentos de modelos de negócios após F&As sugeriram uma dependência em relação ao modelo subjacente das companhias envolvidas, em particular daquela que é adquirida.

## Agradecimentos

## Referências